\documentclass[aps,prb,twocolumn,showpacs,citeautoscript,reprint,longbibliography]{revtex4-2}
\usepackage{boldline,multirow}
\usepackage{float}
\usepackage{graphicx,natbib}
\usepackage{bm,amsmath,amssymb,wasysym,amsfonts,dcolumn}
\usepackage[colorlinks=true, urlcolor=blue, linkcolor=blue, citecolor=blue, pdfborder={0 0 0}]{hyperref}
\usepackage{ulem}
\usepackage[usenames,dvipsnames]{xcolor}
\usepackage{lipsum}
\usepackage{epstopdf}
\usepackage{cleveref}
\crefname{figure}{Fig.}{Figs}
\crefname{table}{Table}{Tables}
\crefname{section}{Sec.}{Sections}

\newcommand{\ket}[1]{\left| #1 \right>} 
\newcommand{\bra}[1]{\left< #1 \right|} 

\begin{document}

\title{Fully resolved currents from quantum transport calculations}

\author{R.S.Nair}
\date{\today}
\affiliation{Faculty of Science and Technology and MESA$^+$ Institute for Nanotechnology, University of Twente, P.O. Box 217,
		7500 AE Enschede, The Netherlands}
\author{P.J.Kelly}
\affiliation{Faculty of Science and Technology and MESA$^+$ Institute for Nanotechnology, University of Twente, P.O. Box 217,
	7500 AE Enschede, The Netherlands}
	
\begin{abstract}
We extract local current distributions from interatomic currents calculated using a fully relativistic quantum mechanical scattering formalism by interpolation onto a three-dimensional grid. The method is illustrated with calculations for Pt$|$Ir and Pt$|$Au multilayers as well as for thin films of Pt and Au that include temperature-dependent lattice disorder. 
The current flow is studied in the ``classical'' and ``Knudsen'' limits determined by the sample thickness relative to the mean free path $\lambda$, introducing current streamlines to visualize the results. For periodic multilayers, our results in the classical limit reveal that transport inside a metal can be described using a single value of resistivity $\rho$ combined with a linear variation of $\rho$ at the interface while the Knudsen limit indicates a strong spatial dependence of $\rho$ inside a metal and an anomalous dip of the current density at the interface which is accentuated in a region where transient shunting persists.
\end{abstract}

\maketitle

\section{Introduction}
\label{intro}

\begin{figure}[!b]
\includegraphics[width=8.4cm]{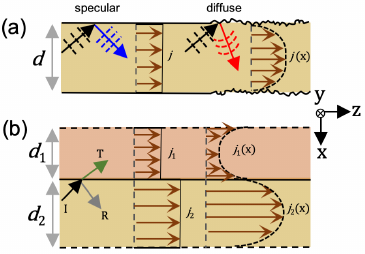} 
\caption{Schematics of (a) electron scattering in a thin film geometry of finite thickness $d$ illustrating the role of surface scattering. An electron wave-packet (depicted by the black arrows) undergoing only specular scattering at the surface (blue arrow) results in a uniform current distribution $j$ flowing in the $z$ direction whereas diffuse scattering from a rough surface (red arrow) results in a non-uniform current profile $j(x)$. (b) Scattering at an interface (thick black horizontal line)  between two slabs of finite thicknesses $d_1$ and $d_2$ where the horizontal dashed lines represent a surface or the next interface in a bilayer/multilayer geometry. At an interface, a part of the incident electron wave (I) is transmitted (T) and the rest is reflected (R) specularly or diffusely eventually leading to current equilibration along $y$. A parallel resistivity model describes $x$-independent resistivities and, corresponding to these, uniform current distributions. A realistic current distribution $j_1(x)$ and $j_2(x)$ resulting from a resistivity gradient across the interface and from details of the scattering (specular and diffuse) is sketched on the right. Note that the spatial dependence is only expected to be significant over a length scale determined by the mean free path of the materials. Coordinate axes are shown for reference.
}
\label{scattering}
\end{figure}

The standard way to measure a bulk resistivity $\rho$ is the four-point-probe technique \cite{Valdes:pire54, vanderPauw:prr58, Miccoli:jpcm15} which assumes isotropic current propagation. In the ongoing pursuit of miniaturization in electronics, such measurements have been extended to study the enhancement of resistivity in thin films \cite{Fuchs:pcps38, Chopra:jap63, Mayadas:apl69, deVries:tsf88, Wu:apl04, Chawla:apl09, Chawla:prb11, Dutta:jap17} (and wires \cite{Steinhogl:prb02, Josell:armr09, Chawla:prb11}) with a common theme being the estimation of a single effective value of $\rho$ for a given thickness (radius) $d$ \cite{Josell:armr09}. When the Fermi wavelength of the conduction electrons is comparable to $d$, the wave nature of electrons gives rise to finite size effects \cite{Datta:95} that must be taken into consideration. If the mean-free-path $\lambda$ is comparable to $d$, the whole concept of a local resistivity becomes moot and as illustrated in \cref{scattering}(a), specular and diffusive reflection from surfaces play a role in determining the current distribution in such films \cite{Fuchs:pcps38, Sondheimer:ap52}. The corresponding case of a multilayer is illustrated in \cref{scattering}(b) where interfaces give rise to specular and diffusive scattering and the finite transmission through interfaces leads to shunting of current which is typically addressed in experiments using parallel resistivity models \cite{Barnas:prb90, Liu:arXiv11}.

The field of spintronics originated with multilayers comprising alternating thin films of magnetic and nonmagnetic metals \cite{Baibich:prl88, Binasch:prb89} and these continue to play a pivotal role \cite{Nagaosa:rmp10, Hoffmann:ieeem13, Sinova:rmp15, Hoffmann:prap15}. The accurate estimation of various spin transport parameters is intimately connected with knowing how much charge current flows in the different layers that are of the order of 1-10 nm thick. A very recent attempt to determine this current distribution combined different thin film resistivity models with four-point-probe measurements for a large number of samples where the individual layer thicknesses were varied systematically \cite{Stejskal:prb20}. This indirect approach was made necessary by the absence of a direct method to observe how current flows in the different layers of multilayer samples. Stejskal et al. concluded their study by emphasizing the need for more detailed structural characterization in order to be able to reduce the non-negligible variations they found in the model-dependent allocation of currents to individual layers. 

Although the transport properties of metals are known to be dominated by states close to the Fermi energy \cite{Ziman:60, Allen:96, Savrasov:prb96b}, there have been few attempts to include the full complexities of the Fermi surfaces associated with partially filled $d$ bands in theoretical studies of transport parallel to the surfaces of thin films \cite{ThinFilmTransport}, or in the context of multilayers, parallel to the interfaces, the so-called current-in-plane (CIP) configuration \cite{Camblong:prl92, Zhang:prb92, Zhang:prb95}. The most sophisticated model used by Stejskal et al. \cite{Stejskal:prb20} was the phenomenological electron gas model of Fuchs \cite{Fuchs:pcps38} and Sondheimer \cite{Sondheimer:ap52}, generalized to treat two different surfaces \cite{Lucas:jap65} and to include transmission through interfaces between two different metals \cite{Barnas:prb90} as well as the effect of grain boundaries \cite{Mayadas:apl69, Mayadas:prb70}. 
The purpose of this paper is therefore to explore the possibility of calculating the spatial distribution of currents in realistic multilayers and thin films entirely from first principles including temperature-induced lattice disorder \cite{LiuY:prb11, *LiuY:prb15}.
To do so, we introduce a discrete scheme to interpolate local currents \cite{Wesselink:prb19} calculated using a fully relativistic DFT based scattering code \cite{Starikov:prb18} and apply it to evaluate the full spatial profile of currents in thin films of Pt and Au which are of interest to the spintronics community as well as in Pt$|$Au and Pt$|$Ir multilayers. 

The paper is arranged as follows. In \cref{QT} some aspects of the scattering problem that are relevant for the calculation of interatomic currents are briefly summarized. The planar averaging introduced in Ref.~\onlinecite{Wesselink:prb19} is relaxed in \cref{Int} to obtain fully spatially resolved local currents on a three dimensional grid. Inspired by fluid physics we introduce streamlines to visualize the current flow in \cref{streamlines}. Although the same methodology can be straightforwardly applied to study the spatial distribution of spin currents, the present work will for simplicity focus on charge currents. In \cref{results} the methodology presented in the previous section is illustrated: in the Knudsen limit for a thin film of Au and a Pt$|$Au multilayer in \cref{Kundsen}; in the classical diffusive limit for a thin Pt film and a Pt$|$Ir multilayer in \cref{Class}. The non-negligible effect of the choice of lead material is examined in \cref{Leads} and we conclude with a brief discussion and outlook in \cref{discussion}.
  
\section{Methods}
\label{method}

\subsection{Quantum transport}
\label{QT}

\begin{figure}[t]
\includegraphics[width=8.4cm]{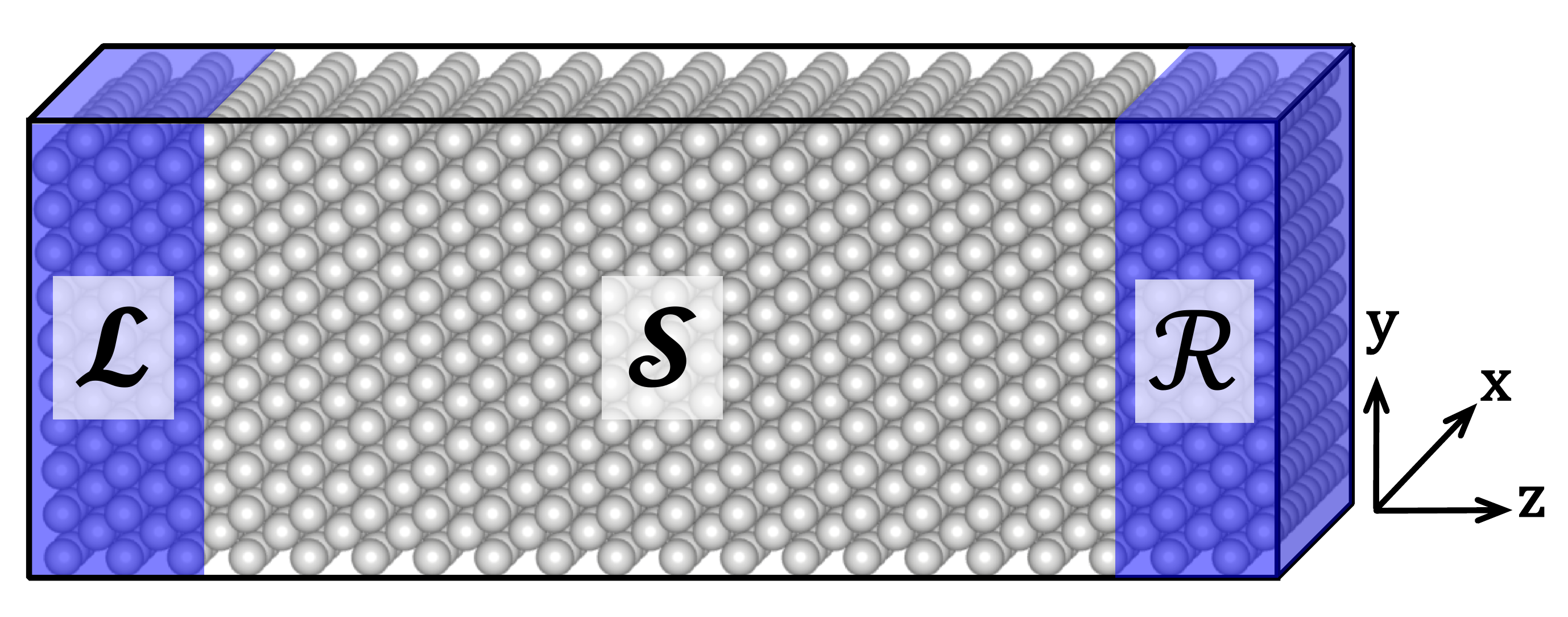} 
\caption{A scattering region ($\mathcal{S}$) is sandwiched between lattice-matched ballistic left ($\mathcal{L}$) and right ($\mathcal{R}$) leads which are semi-infinite in the $\pm z$-direction respectively. Superlattice periodicity is imposed in the $xy$ plane by means of an $N \times M$ supercell in the $x$ and $y$ directions, respectively. This construction makes it possible to simulate a wide range of disordered systems. Coordinate axes are shown for reference for an fcc lattice with $N=M=10$ and $x$=[110], $y$=[1$\bar{1}$0], $z$=[001]. 
}
\label{schematic}
\end{figure}

A typical two-terminal transport configuration is sketched in \cref{schematic} with a scattering region ($\mathcal{S}$) sandwiched between ideal left ($\mathcal{L}$) and right ($\mathcal{R}$) crystalline leads. In the adiabatic approximation, atoms in the scattering region are displaced from their mean positions with a Gaussian distribution of displacements characterized by a root-mean square displacement $\Delta(T)$ chosen to reproduce the experimental resistivity \cite{HCP90} at a given temperature $T$ \cite{LiuY:prb11, *LiuY:prb15}. Such disorder would break the translational symmetry completely and make it impossible to solve the Schr{\"{o}}dinger equation. To remedy this, we introduce periodic boundary conditions in the $xy$ plane with an $N \times M$ ``lateral supercell'' comprising $N$ and $M$ unit cells in the $x$ and $y$ directions, respectively, whereby the disorder is assumed to be periodic. It turns out that remarkably small supercells are sufficient to eliminate observable effects of the residual periodicity as long as the temperature is not too low \cite{Starikov:prb18}.

 The transport problem now reduces to one of solving the single particle Schr{\"{o}}dinger equation inside region $\mathcal{S}$ using the propagating Bloch states of the periodic semi-infinite leads as boundary conditions. To do so in practice, we make use of a ``wave function matching'' (WFM) scheme \cite{Ando:prb91, Khomyakov:prb05, Zwierzycki:pssb08} formulated \cite{Xia:prb06, Starikov:prb18} for a basis of tight binding (TB) muffin tin orbitals (TB-MTO) \cite{Andersen:prl84, Andersen:85, Andersen:prb86} and the atomic spheres approximation (ASA) \cite{Andersen:prb75}. TB-MTOs form a localized orbital basis $|i\rangle$ with $i=Rlms$ where $R$ is an atom site index and $lms$ have their conventional meaning. In terms of the basis $|i\rangle$, the wavefunction $\Psi$ can be expressed as
\begin{equation}
\label{eq:psi}
|\Psi\rangle= \sum_i |i\rangle \langle i|\Psi\rangle
\end{equation}
and the Schr\"odinger equation becomes a matrix equation with matrix elements $\langle i |H| j \rangle$. $\Psi$ is a vector of coefficients with elements $\psi_i \equiv \langle i|\Psi\rangle $ extending over all sites $R$ and over the orbitals on those sites, for convenience collectively labelled as $i_R$. $| \Psi_R \rangle$ is a projection of the total wave function $|\Psi \rangle$ onto the orbitals on atom $R$ 
\begin{equation}
   | \Psi_R \rangle = \sum_{i_R} \left|i_R \rangle \langle i_R \right| \Psi \rangle \,.
\end{equation}
The minimal TB-MTO basis along with the local density approximation (LDA) of density functional theory (DFT) \cite{Hohenberg:pr64, Kohn:pr65} makes the scattering problem tractable for scattering regions comtaining $10^4$-$10^5$ atoms. A detailed  description of the TB-MTO-WFM transport scheme can be found in references \cite{Xia:prb06} and \cite{Starikov:prb18}.

\subsection{Interpolation of interatomic currents onto a three dimensional grid}
\label{Int}

We begin with expressions \cite{Wesselink:prb19} for the charge current $j_c^{PQ}$ and spin current $j_{s\alpha}^{PQ}$ between atoms $P$ and $Q$ 
\begin{align}
j_c^{PQ}&=\frac{1}{i\hbar}
    \big[\bra{\Psi_P}H_{PQ}\ket{\Psi_Q} -\bra{\Psi_Q}H_{QP}\ket{\Psi_P} \big] \\
j_{s\alpha}^{PQ}&=\frac{1}{i\hbar}
\big[\bra{\Psi_P}\sigma_{\alpha} H_{PQ}\ket{\Psi_Q} -\bra{\Psi_Q}H_{QP}\sigma_{\alpha}\ket{\Psi_P} \big]
\label{current1}
\end{align}
that are given in terms of the block $H_{PQ}$ of Hamiltonian matrix elements and vectors of expansion coefficients $\langle i_P | \Psi \rangle $ and $\langle i_Q | \Psi \rangle $ obtained by solving the scattering problem. A summation over $lms$ is implicit. ${\bm \sigma}$ is a vector of  Pauli spin matrices $\sigma_{\alpha}$ and $\alpha$ labels the polarization direction of the spin current. 
The materials whose transport properties we wish to study are crystalline materials or substitutional alloys at finite temperatures whose constituent atoms are displaced at random from the sites of a Bravais lattice. Determining the spatial distribution of currents in the scattering region thus requires interpolating all $j^{PQ}$ onto regular real space meshes as a function of $x, y, z$.

\begin{figure}[t]
\includegraphics[width=9cm]{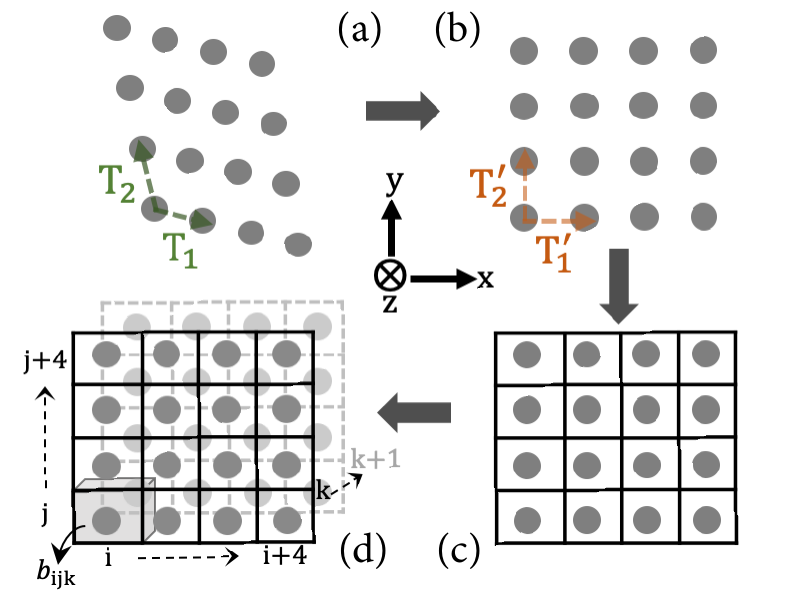} 
\caption{(a) An arbitrary lattice described by the translation vectors ${\bf T}_1, {\bf T}_2$ (and ${\bf T}_3$; not shown) is affine transformed into an equivalent orthogonal lattice (b) described by ${\bf T}'_1, {\bf T}'_2$ (and ${\bf T}'_3$; not shown). (c) The lattice is then discretized into $n_x n_yn_z$ boxes equal to the number of atoms. Here $n_i$ with $i=x,y,z$ is the number of atomic layers in the $i$th direction. 
(d) The final grid where a single box $b_{ijk}\equiv b(x_i,y_j,z_k)$ is explicitly shown in 3D to help visualize the grid. The example shown is for a $4\times 4$ lateral supercell in the $xy$ plane with only two consecutive layers in the $z$ direction shown for clarity.  Cartesian coordinate axes are shown for reference.}
\label{transform}
\end{figure}

In \cref{transform} we illustrate the discretization of an arbitrary transport geometry. To generalize the interpolation and averaging of currents for an arbitrary geometry generated by translation vectors ${\bf T}_1, {\bf T}_2$ and ${\bf T}_3$ that are not necessarily orthogonal to each other, we first perform an affine transformation $\mathcal{T}$ \cite{Modenov:65, Ahuja:ibmsj68, Bamberg:91, Comninos:06} of the translation vectors with a combination of shearing and scaling transformations into a dual space of orthonormal vectors ${\bf T}'_1, {\bf T}'_2$ and ${\bf T}'_3$ that lie along the Cartesian $x, y$ and $z$ coordinate axes. Mathematically, we can express this as
\begin{equation}
\mathcal{T}:\mathcal{R}\rightarrow \mathcal{O}
\end{equation}
where $\mathcal{R}$ is the parent coordinate space and $\mathcal{O}$ is the dual space. Since the collinearity of points along a given direction in the parent space is mapped into a corresponding collinearity in the affine transformed space (\S23 of \cite{Modenov:65}), averaging 
of local quantities in  $\mathcal{O}$ say $x(\equiv {\bf T}'_1)$ can be treated as being equivalent to averaging in the direction of the corresponding translation vector (${\bf T}_1$) in $\mathcal{R}$. We now apply $\mathcal{T}$ to the disordered geometry and map all atomic coordinates from $\mathcal{R}$ to $\mathcal{O}$ thus transforming the disorder from the parent space.
The disordered geometry in $\mathcal{O}$ is then divided into boxes whose dimensions $D_x$, $D_y$ and $D_z$ are determined from the average distance between consecutive atomic layers in the $x$, $y$ and $z$ directions, respectively, such that the number of boxes is equal to the number of atoms and each box contains exactly one atom; the latter is guaranteed if the temperature-induced atomic displacements are much less than the interatomic separations. The regular lattice of boxes is constructed in such a way that the centres of gravity of the atomic coordinates and boxes coincide.


\begin{figure}[t]
\includegraphics[width=8.8cm]{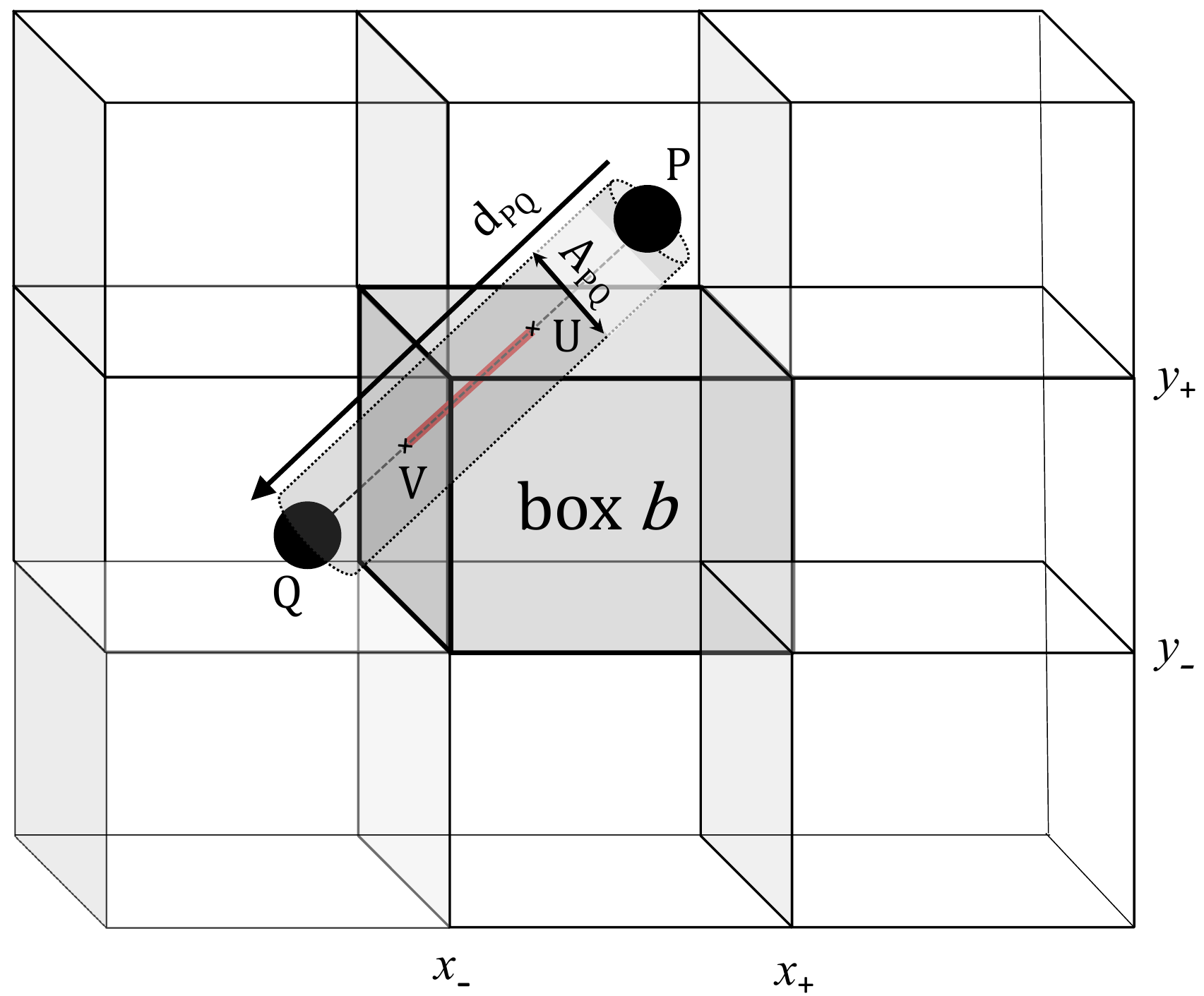} 
\caption{Illustration of the discrete current scheme where the grid is zoomed to show how the current $j^{PQ}$ between atoms $P$ and $Q$ is interpolated into the box $b$. The current contribution from ``wire'' $PQ$ to the box $b$ comes from the segment $UV$ (coloured red) that lies inside the box.
}
\label{fig:box}
\end{figure}

${\bf j}^{PQ} \equiv (j_c^{PQ},j_{sx}^{PQ},j_{sy}^{PQ},j_{sz}^{PQ})$ is imagined as a current through a wire connecting the positions of atoms $P$ and $Q$ with (arbitrary) cross section $A_{PQ}$, see \cref{fig:box}. Since microscopic details of the spatial distribution of ${\bf j}^{PQ}$ are unknowable, we assume a homogeneous flux of current between $P$ and $Q$. The tensor current density of the wire is such that $\overleftrightarrow{j}^{PQ}V_{PQ}={\bf j}^{PQ}\otimes{\bf d}_{PQ} $ where $V_{PQ}=A_{PQ}d_{PQ}$ is the volume of the wire $PQ$, ${\bf d}_{PQ}$ is the vector pointing from $P$ to $Q$ and $d_{PQ}$ its length. The direct product ${\bf j}^{PQ}\otimes{\bf d}_{PQ}$ is estimated in the parent space $\mathcal{R}$ as it depends on the components of ${\bf j}^{PQ}$ and ${\bf d}_{PQ}$. Since the affine transformation $\mathcal{T}$ preserves ratios of distances between points in a line  (\S24 of \cite{Modenov:65}), the current contribution to each box can be determined in the dual space $\mathcal{O}$ using a linear interpolation scheme.

Unlike the planar averaged scheme \cite{Wesselink:prb19} where a one-dimensional interpolation was enough to evaluate the variation in the $z$ direction only, a three dimensional interpolation is implemented here. A general case is shown in \cref{fig:box} for a pair of atoms $P$ and $Q$ in $\mathcal{O}$ with their centres at $(x_P, y_P, z_P)$ and $(x_Q, y_Q, z_Q)$ respectively. Note that $P$ and $Q$ can be inside or outside the box ``$b$''. Only boxes that are intersected by the wire $PQ$ receive a contribution from the current ${\bf j}^{PQ} $ which makes it a classic computational problem of ``collision detection'' where one seeks the point of intersection of the path of an object and a surface of interest. Mathematically, this requires simultaneously solving the equation of line $PQ$ connecting the pair of atoms $\{P,Q\}$ with equations describing the six faces of each box for all possible $\{P,Q\}$. This amounts to $7\times C(n,2) \times C(n,1)=\frac{7}{2}(n^3-n^2)$ equations where $n$ is the number of atoms in the geometry with a computational effort that scales as $\sim O(n^9)$. To efficiently perform interpolations for systems with multiple configurations of $10^4-10^5$ atoms, we instead take advantage of the orthogonality of the dual space $\mathcal{O}$ and employ a line clipping algorithm \cite{Liang:acmtg84} to determine all boxes which are intersected by wire $PQ$ as well as the points of intersection $U$ and $V$. 

The line $PQ$ is first parametrized as
\begin{equation}
(x, y, z)=(x_P, y_P, z_P) - c(x_P-x_Q, y_P-y_Q, z_P-z_Q) 
\end{equation}
for $0\le c\le 1$. Each box can be described by six boundaries, two for each direction: $\{x_-$, $x_+\}$, $\{y_-$, $y_+\}$, $\{z_-$, $z_+\}$, allowing us to write six inequalities for $P\ne Q$
\begin{equation}
\frac{\alpha_P-\alpha_+}{\alpha_P-\alpha_Q} 
\le c \le 
\frac{\alpha_P-\alpha_-}{\alpha_P-\alpha_Q} \hspace{1em} \text{for}\hspace{1em}\alpha=x,y,z .
\label{t_parameter}
\end{equation}
Equation~\eqref{t_parameter} is only satisfied by the points on the line $PQ$ that lies inside the box. Then $c$ takes a range of continuous values from which the intercepts $U$ and $V$ can be obtained as 
\begin{subequations}
\begin{align}
U&=(x_P, y_P, z_P) -\min(c)(x_P-x_Q, y_P-y_Q, z_P-z_Q)  \\ 
V&=(x_P, y_P, z_P) -\max(c)(x_P-x_Q, y_P-y_Q, z_P-z_Q)
\end{align}
\end{subequations}
Note that $U=(x_P, y_P, z_P)$  when $P$ lies inside $b$  and $V=(x_Q, y_Q, z_Q)$ when $Q$ lies inside $b$, which we denote as  $P \in b$, $ Q \in b$, respectively in the following.
 We define a parameter $\beta$ that indicates how much of the wire lies outside the box on either side. 
\begin{subequations} 
\begin{align}
\beta_{QP,b} &= \begin{cases}
0                     & {\rm if \,} Q \in b \\
\frac{d_{QV}}{d_{QP}} & {\rm if \,} Q \notin b
\end{cases}\\
\beta_{PQ,b} &=\begin{cases}
0                     & {\rm if \,} P \in b \\
\frac{d_{PU}}{d_{PQ}} & {\rm if \,} P \notin b
\end{cases}
\end{align}
\end{subequations}
where $d_{AB}$ is the norm of the vector ${\bf d}_{AB}$ pointing from point $B$ to point $A$. (In the planar averaged scheme \cite{Wesselink:prb19}, only the projection $d^z_{AB}$ of ${\bf d}_{AB}$ on $z$, is used to evaluate $\beta_{PQ}$.) Since the current changes between $Q$ and $P$, we make a linear interpolation 
\begin{equation}
{\bf j}^{PQ}(c) = c \,{\bf j}^{PQ}-(1-c){\bf j}^{QP}
\end{equation}
Thus, for a box of volume $V_b$, the contribution from $\overleftrightarrow{j}^{PQ}V_{PQ}$ is given by 
\begin{eqnarray}
\int^{1-\beta_{PQ,b}}_{\beta_{QP,b}} &&{\bf j}^{PQ}(c)\otimes {\bf d}_{PQ} \, dc  = \nonumber \\
\tfrac{1}{2} &&\left[ \left( 1-\beta_{PQ,b}\right)^2 - \left(\beta_{QP,b}\right)^2\ \right] {\bf j}^{PQ}\otimes {\bf d}_{PQ} \nonumber \\
+\tfrac{1}{2} &&\left[ \left(1-\beta_{QP,b}\right)^2 - \left(\beta_{PQ,b}\right)^2\ \right]{\bf j}^{QP}\otimes {\bf d}_{QP} \,.
\end{eqnarray}
Note that ${\bf d}_{PQ}={\bf-d}_{QP}$. The average current density tensor in the box $b$ is then
\begin{equation}
\stackrel{\leftrightarrow}{j_b} =
\frac{1}{V_b} \sum_{P,Q} \tfrac{1}{2} \left[ \left( 1-\beta_{PQ,b}\right)^2 - \left(\beta_{QP,b}\right)^2\ \right] {\bf j}^{PQ}\otimes {\bf d}_{PQ} 
\end{equation}
and we take this value of $\stackrel{\leftrightarrow}{j_b}$ to be the average current density at the centroid of $b$. By interpolating all interatomic currents into all boxes $b$, we obtain the complete spatial variation of the current density. Multiplying the current density by the cross sectional area of the box $b$ perpendicular to $z$ yields the current per unit voltage applied across the leads, a conductance. Summation of the normalized currents for all boxes lying in a given $xy$ plane should then be equal to the Landauer-Buttiker conductance that is calculated independently of the interatomic currents and interpolation scheme. This  provides a check of the whole local current formalism. Finally, centroids of the boxes are affine transformed back into the parent space,
\begin{equation}
\mathcal{T}':\mathcal{O}\rightarrow\mathcal{R} .
\end{equation}

We observe spatial oscillations in calculated spin currents because of the interference between reflected and incident electron matter waves. Although these oscillations are real, they are not present in semiclassical descriptions of transport. Because they are found to be attenuated away from the leads in parallel with the corresponding decrease in the unscreened particle accumulation, we follow Ref.~\onlinecite{Wesselink:prb19} and use the latter to reduce these quantum fluctuations to facilitate analysis using semiclassical transport formulations. Since lateral supercell sizes do not exceed more than a few hundred atoms in our calculations, we perform such averaging using only the planar averaged unscreened particle accumulation.  Details of this averaging can be found in Ref.~\onlinecite{Wesselink:prb19}.

\subsection{Current streamlines}
\label{streamlines}

From here on we only consider the charge current and will therefore drop the subscript $c$. Because of the assumed superlattice periodicity, the average current in the $y$ direction, $\overline{j}_y(x,z)=0$. The visualization of ${\bf j}(\bf r)$ reduces to a problem in two dimensions if we average over $y$ and, to do so, we introduce a current {\it stream function} $\psi(x,z)$ (not to be confused with the wavefunction) by analogy with the velocity stream function in fluid physics \cite{Panton:13}. In the steady state, charge conservation requires that $\nabla.{\bf j}=0$ and for the $xz$ plane, this reduces to
\begin{equation}
\frac{\partial j_x}{\partial x}+\frac{\partial j_z}{\partial z}=0.
\label{2d_divj_c}
\end{equation}
Defining $\psi(x,z)$ such that
\begin{equation}
j_x =\frac{\partial \psi}{\partial z} \;\;\; {\rm and} \;\;\;
j_z =-\frac{\partial \psi}{\partial x}
\label{psi_def}
\end{equation}
automatically leads to \eqref{2d_divj_c} being satisfied. $\psi(x,z)= {\rm constant}$ is a path whose tangent at any point gives the direction of the current vector ${\bf j}$ $=j_x{\bf i}+j_z{\bf k}$ at that point. This defines a {\it streamline} and the volume flow per unit width between streamlines connecting the left ($\mathcal{L}$) and right ($\mathcal{R}$) leads is
\begin{equation}
\psi_{\mathcal{R}} - \psi_{\mathcal{L}} = \int_{\mathcal{L}}^{\mathcal{R}} d\psi = \int_{\mathcal{L}}^{\mathcal{R}} \big[j_x dz - j_z dx \big]
\end{equation}
This region can be thought of as a conducting strip carrying a constant flux of current analogous to a streamtube for incompressible fluid flow such that crowding of streamlines at a region in the flow-field indicates a local increase in the magnitude of the current \cite{Panton:13}.

\subsection{Mean Free Path}
\label{MFP}

In the relaxation time approximation (RTA), the conductivity is given in terms of the ${\bf k}$ dependent velocities ${\bm \upsilon({\bf k})}=\frac{1}{\hbar} \nabla_{\bf k}\varepsilon({\bf k})$ as
\begin{equation}
\label{RTA}
\sigma_{ij}= e^2 \iiint \frac{d^3k}{8\pi^3} \tau({\bf k}) \,
           \upsilon_i({\bf k})\upsilon_j({\bf k}) 
           \Big(-\frac{\partial f}{\partial \varepsilon}\Big)_{\varepsilon=\varepsilon({\bf k})} .  
\end{equation}
In the low temperature limit $-\frac{\partial f}{\partial \varepsilon} \rightarrow \delta(\varepsilon - \varepsilon_F)$ and \eqref{RTA} becomes an integral over the Fermi surface $S_F$. Assuming additionally that $\tau({\bf k})=\tau(\varepsilon({\bf k})) $ then 
\begin{equation}           
\sigma =  e^2 D(\varepsilon_F) \tau(\varepsilon_F) \langle v^2_F \rangle 
\end{equation}
where $D(\varepsilon)$ is the density of states. Both $D(\varepsilon_F)$ and $\langle v^2_F \rangle$ can be evaluated from standard bulk LMTO electronic structure calculations \cite{footnote3} and since $\sigma \equiv 1/\rho$ is known \cite{HCP90, footnote1}, $\tau$ can be expressed as         
\begin{equation}
\tau = \frac{\sigma}{e^2 D(\varepsilon_F) \langle v^2_F \rangle}  
\end{equation}
and the mean free path can be estimated as $\lambda=\tau \langle v^2_F \rangle^{1/2}.$ 

\section{Results}
\label{results}

\begin{figure}[t]
\includegraphics[width=8.8cm]{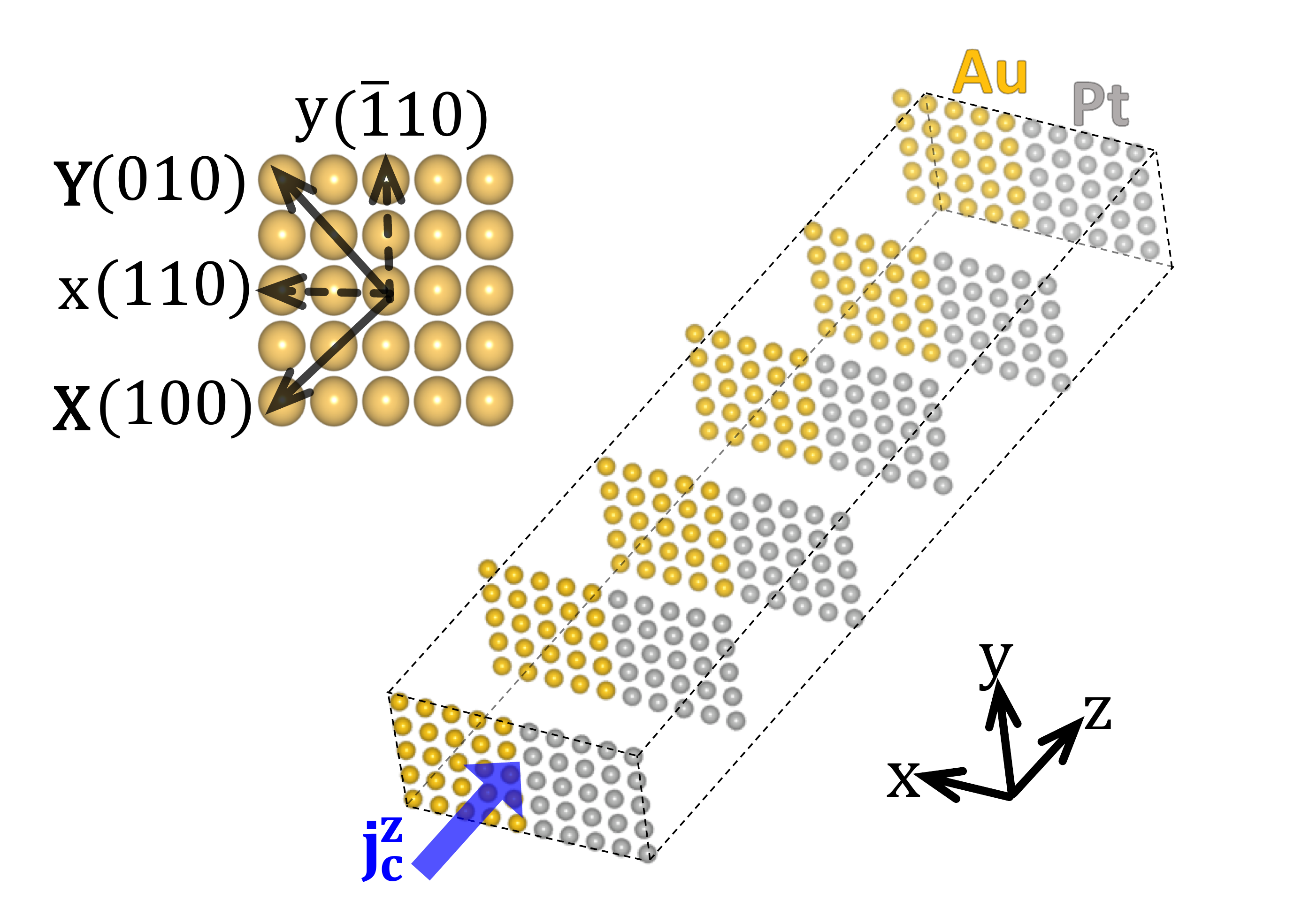}
\caption{Sketch of an $N \times M$ lateral supercell used to model transport in a lattice-matched AB multilayer with A=Au and B=Pt. For clarity, only six layers in the $z$ direction are explicitly shown, the separation of the layers is exaggerated and the leads sandwiching this geometry in the $\pm z$-directions are not shown. The $x$ direction is the crystal [110] direction. Part of an fcc layer perpendicular to the [001] direction with in-plane crystallographic directions is shown on the left. The conventional cubic axes of an fcc lattice are $X, Y$ and $ Z(\equiv z)$.
	}
\label{fig:injgeom}
\end{figure}

Different regimes of electron transport can be identified depending on the ratio of the electron mean free path $\lambda$ to the critical dimension $d$ of the scattering geometry that is the Knudsen number ($\rm Kn$). When $\lambda \ll d$, we are in the classical limit where the flow of current is well described by Ohm's law. The other extreme is the Knudsen limit where $\lambda \gg d$, size effects and interface or surface scattering dominate and transport deviates from Ohm's law \cite{Wexler:pps66}. To illustrate the three-dimensional current scheme presented above, we consider thin films and two-component ...A$|$B$|$A$|$B... multilayers where the thickness of the $i$th layer is $d_i$. In this paper, only fcc metals are considered as sketched schematically in \cref{fig:injgeom} with a charge current flowing in the [001] direction, parallel to the A$|$B interfaces. A k-point sampling of $\sim \frac{160}{N}\times \frac{160}{M}$ for an $N\times M$ supercell is used throughout the paper. From now on we use the terms current and current density interchangeably. Unless stated otherwise, all currents are averaged over $y$ and are calculated for a temperature $T=300\,$K (``room temperature''); when averaging currents over the scattering region, $z \in \mathcal{S}$, a few layers close to the leads where transient effects are observed are omitted. 

\subsection{Knudsen limit}
\label{Kundsen}

\subsubsection{Au thin film}
\label{sss:GTF}

We begin by calculating the current in a free-standing [110] oriented thin film of Au. The thin film is modelled as a Au$|$vacuum ``multilayer'' by alternating 60 atomic layers of Au with five layers of ``empty spheres'' (with nuclear charge $Z=0$ to simulate vacuum \cite{Skriver:prb92a, *Skriver:prb92b, Daalderop:prb94}) in the $x$ direction so that $N=60+5 $. A periodicity of $M=3$ layers in the $y$ direction is imposed and the scattering region is 90 atomic layers thick in the $z$ direction, see \cref{fig:injgeom}. A value of the root-mean square displacement $\Delta$ of the atoms in the scattering region was chosen to reproduce the room temperature bulk resistivity of Au, $\rho^{300}_{\rm Au}=2.3 \pm 0.07 \, \mu \Omega \,$cm \cite{HCP90, rho:Au300K}. The thickness of the slab in the $x$ direction is approximately $\tfrac{1}{2} \lambda^{300}_{\rm Au}$ where $\lambda=34.5 \,$nm was estimated in the RTA as described above. Ballistic Au leads were used to minimize transient effects at the lead$|$scattering region interface, between crystalline and thermally disordered Au. 

\begin{figure}[t]
\includegraphics[width=8.8cm]{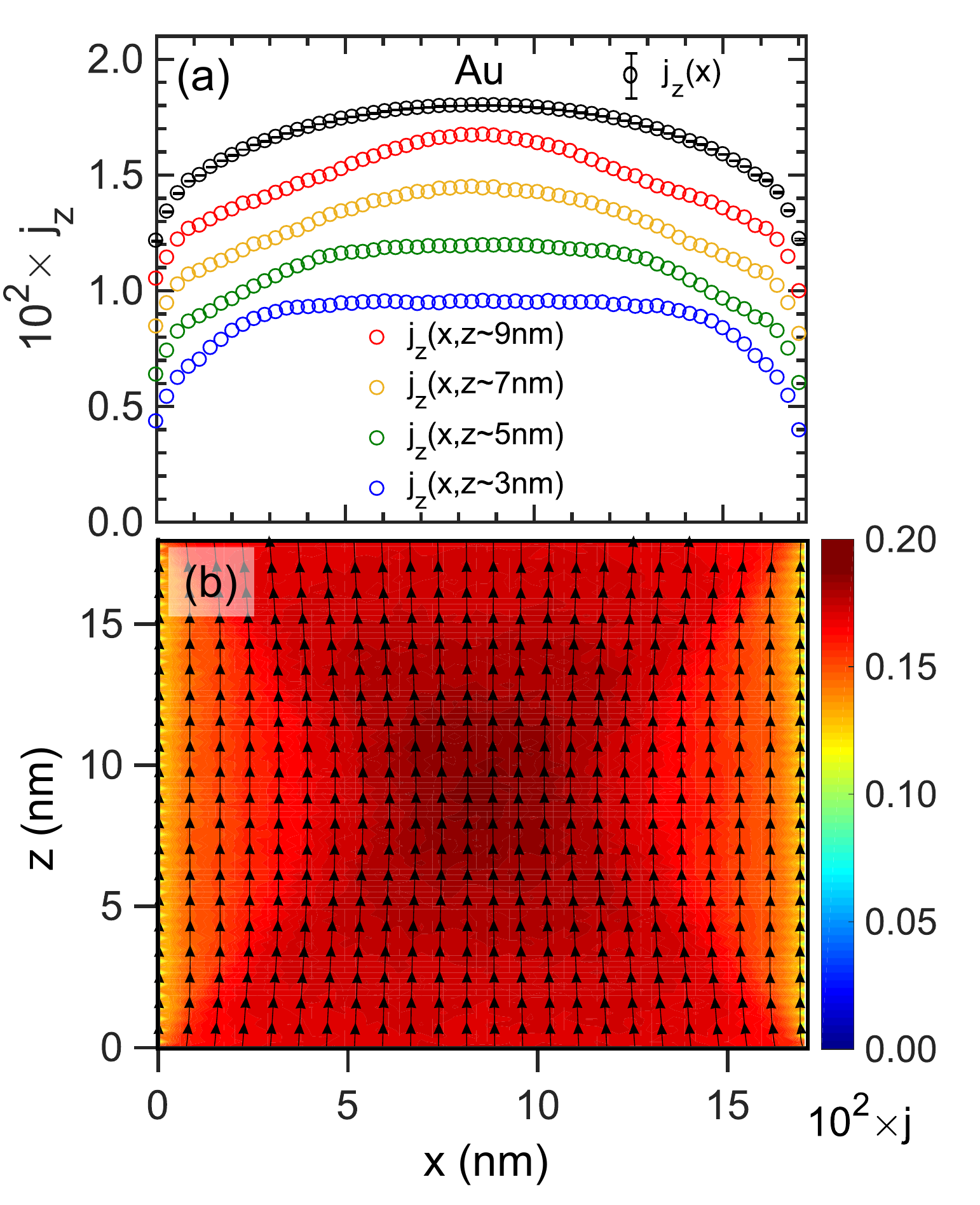}
\caption{Current distribution in a thin film of Au at 335K \cite{rho:Au300K}. (a) Top curve, black symbols: current density $\bar{j}_z(x)$ obtained by averaging over $y$ and $z$. Colour symbols: $\bar{j}_z(x,z_0)$ obtained by averaging over $y$ and $z=z_0\pm5$ layers in the $z$ direction for four different values of $z_0$ that are offset from the black curve in steps of $\Delta j=0.002$ for clarity. The error bars, that are smaller than the symbol size, indicate the average deviation over 10 random configurations of disorder. (b) Streamlines of the current vector ${\bf j}(x,z)=j_x{\bf i}+j_z{\bf k}$ in the $xz$ plane.  The colour contour in the background corresponds to the magnitude of the current. 
}
\label{Au_j}
\end{figure}

In \cref{Au_j}(a) we plot the charge current $\bar{j}_z(x)$ flowing in the $z$ direction averaged over the $y$ and $z$ directions as a function of $x$ (black symbols). This shows a gradual concentration of the current away from the surfaces and towards the middle of the film. A strong $z$ dependence is also apparent from plots of $\bar{j}_z(x,z_0)$ averaged over $y$ and $z=z_0\pm 5$ atomic layers for different values of $z_0 \sim 3, 5, 7, 9 \,$nm measured from the left lead. The current streamlines are plotted in \cref{Au_j}(b) where they are superimposed on a colour map showing the magnitude of the charge current. The colour map shows a larger current density at the centre of Au in both $x$ and $z$ directions. On closer examination, streamlines are not parallel to the $z$ axis but exhibit curvature. This demonstrates that the current distribution has not reached its asymptotic form in $z$ which is not surprising as the length of the scattering region is only about half of the mean free path $\lambda$ \footnote{The relatively short scattering region studied here is a consequence of treating a very large lateral supercell containing $(60+5)\times 90 \times3=17550$ atoms including spin-orbit coupling taking us to the current limits of our computing facilities in terms of both memory and run time}. Apart from a rapid decay of the current density within a few layers of the surface that is described by a ``specularity coefficient'' $p$ in the Fuchs-Sondheimer framework \cite{Fuchs:pcps38, Sondheimer:ap52}, an approximately linear variation of the conductivity from the surface to the middle of the film is observed at the centre of the scattering region furthest from the leads. An ``effective resistivity'' clearly conceals substantial variation in the current density for film thicknesses that are commonly used in spintronics.

\subsubsection{Pt$|$Au multilayer}

\begin{figure*}[t]
\includegraphics[width=\linewidth]{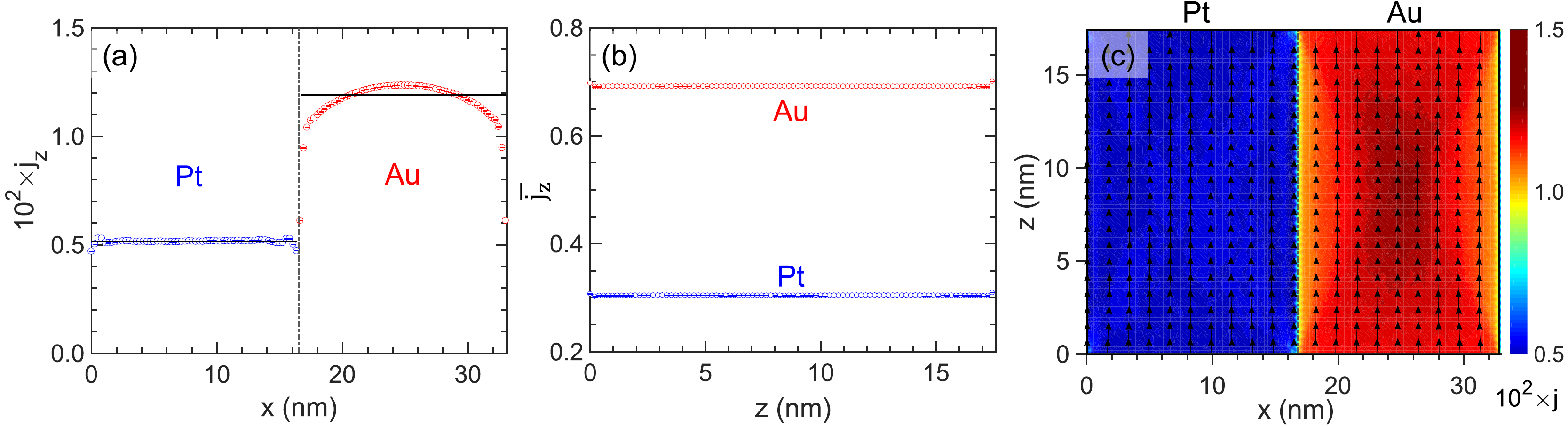}
\caption{Distribution of the charge current $\bar{j}_z$ in a Pt$|$Au multilayer. 
For $\bar{j}_z(x)$ in (a) $j_z(x,y,z)$ is averaged over $y$ and over $z$. The horizontal black lines indicate the mean values obtained by averaging separately over $x \in$(Pt,Au).  
For $\bar{j}_z(z)$ in (b) $j_z(x,y,z)$ is averaged over $x$ and $y$ with $x \in \,$Pt and $x \in \,$Au separately \cite{xyaveraging}.
(c) Streamlines of the current vector ${\bf j}(x,z)=j_x{\bf i}+j_z{\bf k}$ averaged over $y$ and superimposed on a colour contour of $j=\sqrt{j_x^2+j_z^2}$ in the $xz$ plane. Error bars represent the mean deviation over 10 configurations of thermal disorder.}
\label{AuPt_j}
\end{figure*}

Modern experiments frequently use heterostructures in which the thicknesses $d_i$ of constituent layers are comparable to the corresponding bulk mean free paths $\lambda_i$. To demonstrate the deviation from bulk behaviour we choose a Pt$|$Au multilayer. Both Pt and Au are fcc metals with only a $2\%$ lattice mismatch so that such an epitaxial multilayer might be prepared without undue structural disorder. 
Thermal disorder corresponding to the room temperature bulk resistivities of Au, $\rho^{300}_{\rm Au}=2.6 \pm 0.07 \, \mu \Omega \,$cm and Pt, $\rho^{300}_{\rm Pt}=10.8 \pm 0.5 \, \mu \Omega \,$cm, was used with different mean square displacements in the Au and Pt layers. The mean free path in Au, $\lambda^{300}_{\rm Au} = 34.5 \,$nm is almost ten times that in Pt, $\lambda^{300}_{\rm Pt}=3.74 \,$nm \cite{Nair:tbp21a}. Choosing $d_{\rm Au} < \lambda^{300}_{\rm Au}$ and $d_{\rm Pt} > \lambda^{300}_{\rm Pt}$ should make any size effect apparent at room temperature. We construct a scattering geometry as shown schematically in \cref{fig:injgeom} with 60 atomic layers each of Pt and Au in the $x$ direction with a periodicity of 3 layers in the $y$ direction and 90 layers thick in the $z$ direction corresponding to a total of 32400 atoms in the scattering region. A charge current is injected from ballistic Au leads in the $z$ direction. The resulting charge current distribution in the Pt$|$Au multilayer is plotted in \cref{AuPt_j}where the error bars, that are smaller than the symbol sizes, correspond to the uncertainty with which the experimental resistivities are reproduced in our scattering calculations.

 The average shunting $\bar{j}_z^{\rm Au}/\bar{j}_z^{\rm Pt} \sim 2.3$ i.e, the ratio of the mean current value in Au to that in Pt, indicated by the solid black horizontal lines in each layer plotted in \cref{AuPt_j}(a), is much lower than expected from the ratio of the bulk resistivities $\rm\rho^{300}_{Pt}/\rho^{300}_{Au}\sim 4.2$. The charge current is seen to be constant and saturated inside Pt while in Au a rapid increase in the two atomic layers adjacent to the interface followed by a continuous variation to the centre of the layer is observed. We note a small, anomalous dip in the current density in the Pt layers next to the  interface. The current density variation at the interface and in the Au slab is clearly not amenable to description using a single resistivity. An almost immediate saturation of the total current carried by the Au and Pt layers is observed when these are plotted as a function of $z$ in \cref{AuPt_j}(b) making our results independent of the length of the scattering region. The streamlines plotted in \Cref{AuPt_j} are parallel to the $z$ axes inside Pt indicating that there is no net flow of current across the interface into Au as asymptotic shunting of the total charge current density is reached in the $z$ direction. However, a redistribution of the current inside Au is visible from the color contour and small curvature of the streamlines that is similar to what we saw for the Au thin film. As the total current in each layer is independent of $z$, one might use a parallel resistance model to estimate the current shunting noting, however, that this would not correctly describe microscopic details of the variation near the interface and in the Au layer just presented. 
 
\subsection{Classical limit}
\label{Class}
 
The classical diffusive limit is achieved when the mean free path is much shorter than other critical dimensions of the structures being studied, $\lambda \ll d$. Memory requirements currently limit the lateral supercell size to $H=N \times M \approx 300-400$ atoms for which the maximum length of scattering region is about $L \approx 130$ atoms when spin-orbit coupling is included. Because the computational effort for metallic systems scales as approximately $H^2 L$ \cite{Xia:prb06, Starikov:prb18}, considerably longer scattering regions can be studied with smaller lateral supercells. To attain the diffusive limit we need to consider high resistivity materials or study elevated temperatures or both.  

\subsubsection{Pt thin film}
\label{sss:ptf}

We first study charge transport through a thin film of Pt, modelling a free-standing [110] oriented Pt layer as a Pt$|$vacuum multilayer with $N$ layers of Pt and 5 layers of ``empty spheres'' repeated periodically in the $x$ direction. The effect of increasing the default periodicity of three atomic layers in the $y$ direction to five atomic layers is studied. To minimize lead$|$scattering- region interface effects, Pt leads are used. We calculate the average resistivity for different values of $N$ and identify the thickness at which it saturates to the bulk value. At $T=300\,$K, the bulk resistivity of Pt is $\rho^{300}_{\rm Pt}=10.8 \pm 0.5\, \mu \Omega \,$cm and $\lambda^{300}_{\rm Pt} \sim 3.74 \, $nm. 

\begin{figure}[t]
\includegraphics[width=8.8cm]{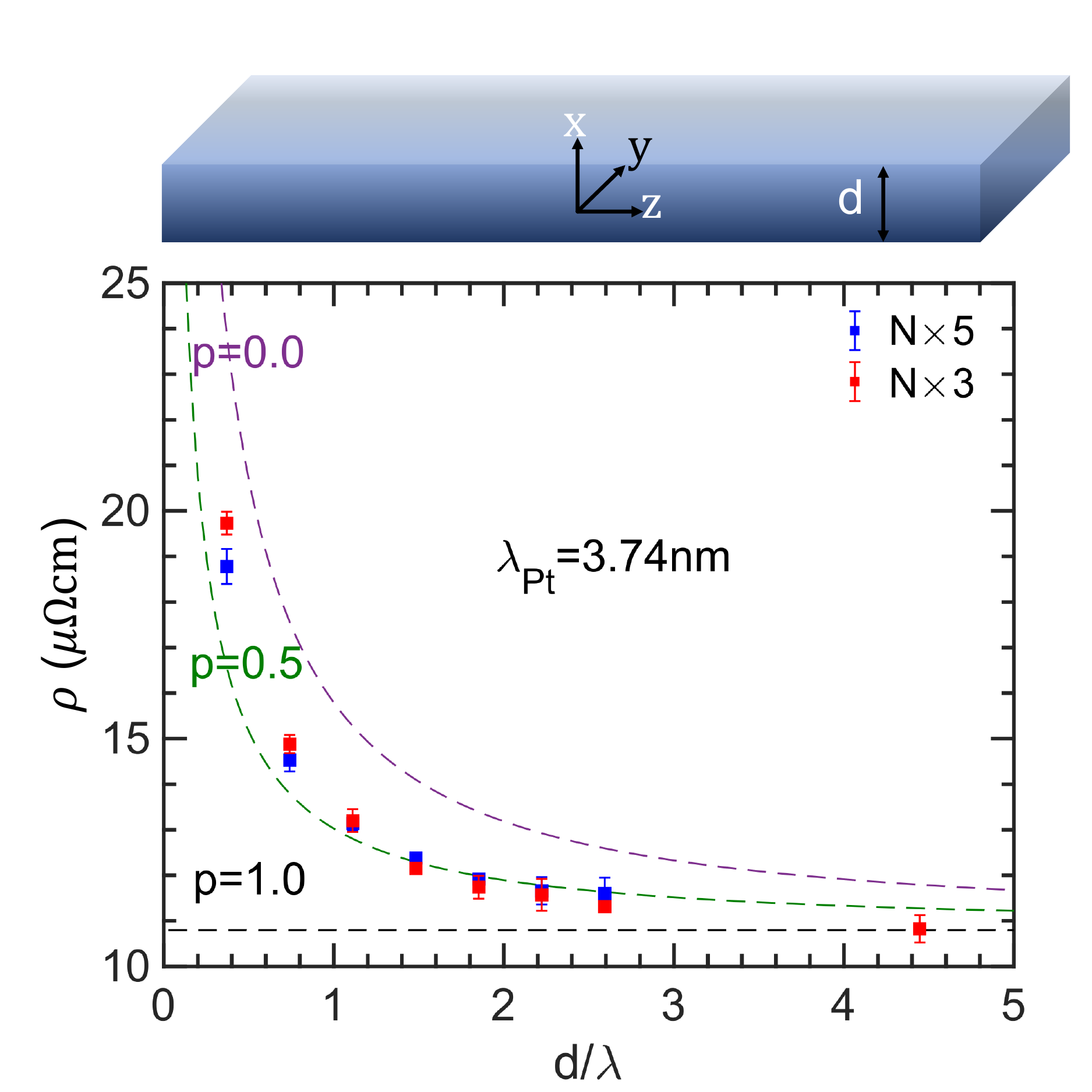} 
\caption{Average resistivity of Pt films calculated as a function of the effective thickness $d/\lambda$ where $d$ is the thickness of the film in the $x$ direction for transport in the $z$(001) direction and $\lambda$ is the mean free path of bulk Pt at RT. The separation of periodically repeated thin films by vacuum (``vacuum thickness'') is modelled using 5 layers of ``empty'' spheres in the $x$ direction. Calculations for each thickness are done for two cases with 5 (blue) and 3 (red) atomic layers repeated periodically in the $y$ direction. The dotted lines are calculated using the Fuchs-Sondheimer model \cite{Fuchs:pcps38, Sondheimer:ap52} for three different values of the  specularity coefficient $p$ that describes the amount of completely diffusive surface scattering: completely specular ($p=1$), partially specular ($p=0.5$) and completely diffusive ($p=0$). According to \eqref{eq:FS}, choosing $p=1$ yields the bulk resistivity $\rho_b$ irrespective of $d/\lambda$.
	}
\label{Pt_res}
\end{figure}

The thickness dependence of the resistivity of a thin film (or wire) is often studied using the ``FS'' model formulated some 70-80 years ago by Fuchs \cite{Fuchs:pcps38} and Sondheimer \cite{Sondheimer:ap52} and named after them, in which surface scattering is treated phenomenologically using Boltzmann transport theory. In the case of a monocrystalline ``free-standing slab'' with only bulk and surface scattering of charge carriers, the thickness dependent resistivity, $\rho$ is given by
\begin{equation}
\begin{split}
\rho(p,d/\lambda) = &\rho_{b}\Bigg[1 - \frac{3}{2(d/\lambda)}(1-p) \\
 &\int_1^{\infty} \!\! \left(\frac{1}{t^3}-\frac{1}{t^5}\right)\frac{1-e^{(d/\lambda)t}}{1-pe^{(d/\lambda)t}}dt \Bigg]^{-1}
\end{split}
\label{eq:FS}
\end{equation} 
where $d$ is the thickness of the slab and $\lambda$ and $\rho_b$ are, respectively, the mean free path and resistivity of the bulk material. The ``specularity coefficient'' $p$ is the fraction of electrons scattered elastically from the surface independent of their velocity and takes values ranging from 1 (specular) to 0 (diffusive). When all the electrons are reflected specularly from the surfaces, the resistivity is identical to that of the bulk; finite-size effects are not considered in the phenomenological FS electron gas model. 

\begin{figure}[t]
\includegraphics[width=8.8cm]{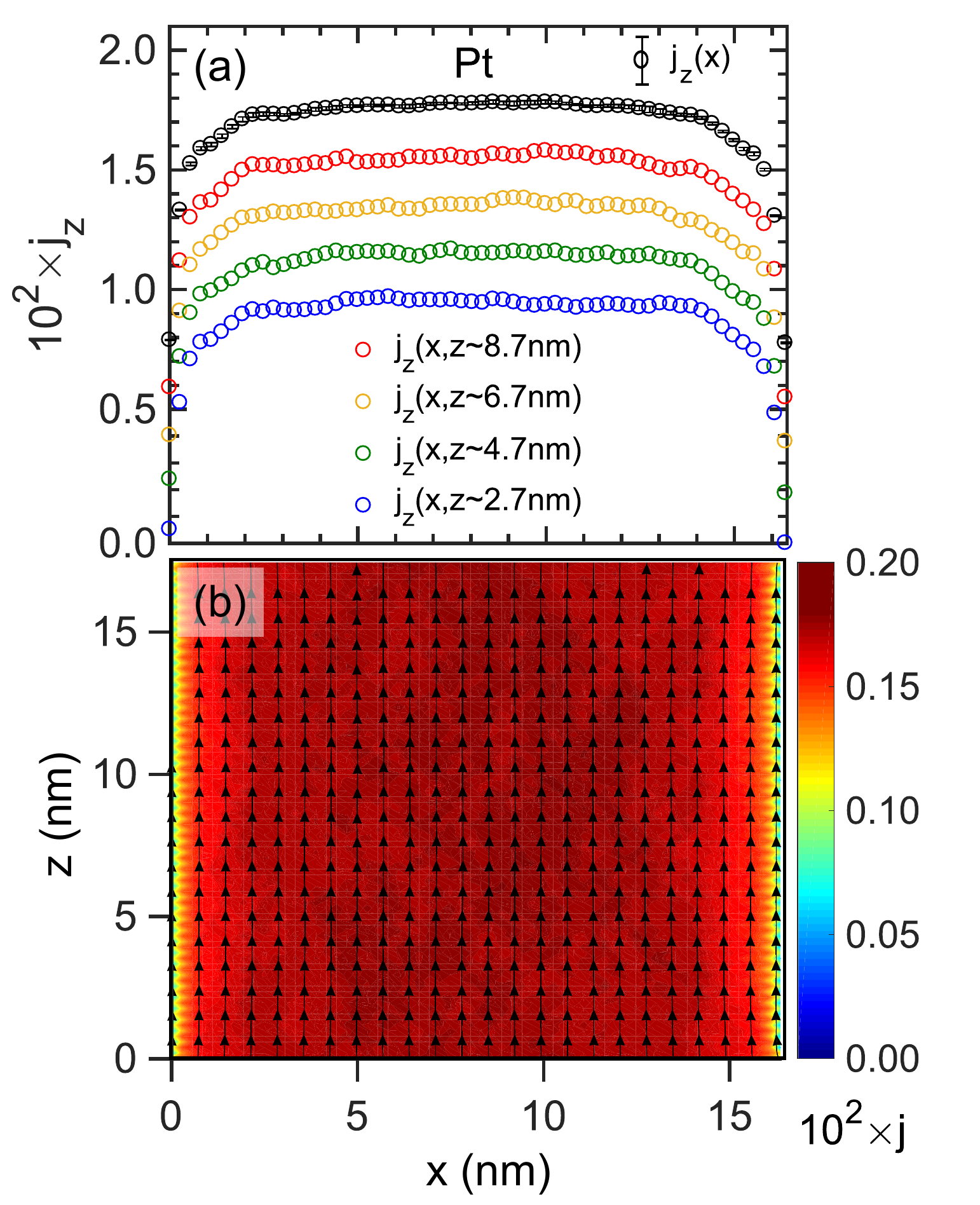}
\caption{(a) Top curve, black symbols: current density $\bar{j}_z(x)$ obtained by averaging over $y$ and $z$. Colour symbols: $\bar{j}_z(x,z_0)$ obtained by averaging over $y$ and $z=z_0\pm5$ layers in the $z$ direction for four different values of $z_0$ that are offset from the black curve in steps of $\Delta j=0.002$ for clarity. The error bars, that are smaller than the symbol size, indicate the average deviation over 10 random configurations of disorder.  (b) Streamlines of the current vector ${\bf j}(x,z)=j_x{\bf i}+j_z{\bf k}$ in the $xz$ plane. The colour contour in the background corresponds to the magnitude of ${\bf j}(x,z)$. 
}
\label{Pt_streamlines}
\end{figure}

\begin{figure*}[ht]
\includegraphics[width=\linewidth]{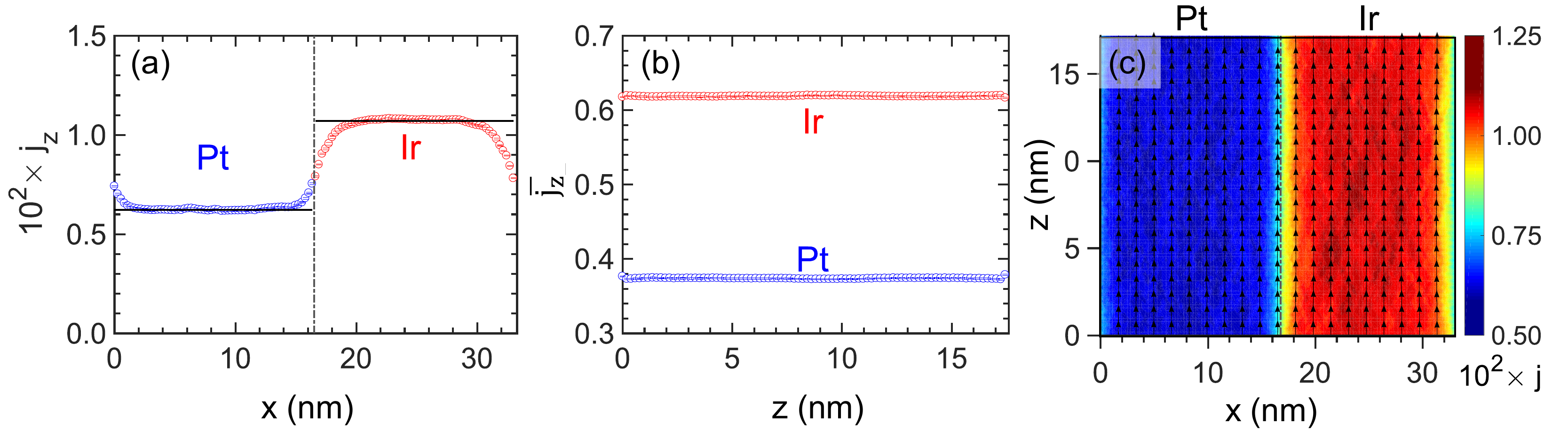}
\caption{Distribution of the charge current $\bar{j}_z$ in a Pt$|$Ir multilayer. For $\bar{j}_z(x)$ in (a) $j_z(x,y,z)$ is averaged over $y$ and over $z$. The horizontal black lines indicate the asymptotic values obtained by averaging separately over $x \in$(Pt,Ir) omitting the region of rapid variation close to the Pt$|$Ir interface in either layer.
		For $\bar{j}_z(z)$ in (b) $j_z(x,y,z)$ is averaged over $x$ and $y$ with $x \in \,$Pt and $x \in \,$Ir separately. (c) Streamlines of the current vector ${\bf j}(x,z)=j_x{\bf i}+j_z{\bf k}$ obtained from averaging $j_z(x,y,z)$ over $y$ and superimposed on a colour contour of $j=\sqrt{j_x^2+j_z^2}$ in the $xz$ plane. The scarcely visible error bars that are smaller than the symbol sizes represent the mean deviation over 10 configurations of thermal disorder.
}
\label{IrPt_j}
\end{figure*}

Values of $\rho$ calculated for different thicknesses of Pt at  $T=300\,$K are plotted in \cref{Pt_res} as a function of $d/\lambda$ where the thin-film enhancement of the resistivity is clear. The resistivity decays to within a few percent of its bulk value when $d \sim 4 \lambda$ and follows the FS model with $p \sim 0.5$ quite well even though the only surface roughness that is included is what results from thermal disorder. The number of layers $M$ in the $y$ direction has been limited to three throughout this paper in order to be able to study films with a reasonably large thickness given by the number of layers $N$ in the $x$ direction.  In \cref{Pt_res} this is shown to be a reasonable compromise for all but the thinnest of films where the results of calculations with $M=3$ and 5 are compared. The thickness dependent resistivity analysis confirms that $d/\lambda$ is the appropriate length scale for transport in slabs with finite thickness. 

For $N=60$ atomic layers corresponding to a film thickness of $\sim 4.4 \lambda^{300}_{\rm Pt}$ and  ``bulk-like'' behaviour, the charge current $\bar{j}_z(x)$ resulting from averaging over $y$ and $z$ is plotted as a function of $x$ in \cref{Pt_streamlines}(a) (black symbols). The current density at different $z$ coordinates (coloured symbols) shows that $j_z$ is only weakly dependent on $z$ unlike what we saw in Au. The reason is that $\lambda^{300}_{\rm Pt}$ is much shorter than the length of the scattering region. In \cref{Pt_streamlines}(b), the current streamlines in the $xz$ plane are superimposed on a colour map of the magnitude $j(x,z) \equiv \sqrt{j_x^2+j_z^2}$ of ${\bf j}(x,z)$ . Within the uncertainties of the calculation the current is constant except very close to the surface where a rapid decay in the current occurs over a length of 2~nm$\,\sim \frac{1}{2}\lambda^{300}_{\rm Pt}$. A more detailed study of transport through thin Pt films will examine the effect of film orientation and surface roughness \cite{Rang:tbp21}.

\subsubsection{Pt$|$Ir multilayer}

We study a Pt$|$Ir multilayer at an elevated temperature of $T=800 \,$K chosen to make it possible to realise ``bulk'' like behaviour with scattering region sizes that are computationally tractable. Lattice disorder was introduced as described in \cref{method} to reproduce the experimental bulk resistivities of Pt, $\rho^{800}_{\rm Pt}=28.1 \pm 0.4 \, \mu \Omega \,$cm and of Ir, $\rho^{800}_{\rm Ir}=16.1  \pm 0.6 \, \mu\Omega \,$cm at 800~K \cite{HCP90} for which $\lambda^{800}_{\rm Pt} \sim1.4 \, $nm and $\lambda^{800}_{\rm Ir} \sim 2.4 \,$nm.

Ir and Pt are both fcc metals with an equilibrium lattice mismatch of only $2\%$. We neglect this mismatch and use a common lattice constant of $a_{\rm Pt}=\rm 0.392 \,$nm in the following. The scattering geometry is constructed as sketched in \cref{fig:injgeom} with 60 (110) planes each of Pt and Ir stacked in the [110] $x$ direction. It consists of 90 (001) atomic layers sandwiched between ballistic Ir leads in the $z$ direction chosen to be the crystal [001] direction with periodicity of three atomic layers in the $y$ direction. A current is injected in the $z$ direction from the Ir leads, parallel to the Pt$|$Ir interface.

 We average $j_z(x,y,z)$ over $y$ and $z$ and plot the resulting $\bar{j}_z(x)$ as a function of $x$ across the interface in \cref{IrPt_j}(a). The horizontal black lines indicate  averaged asymptotic values of current densities calculated separately for $x \in \,$Pt and $x \in \,$Ir. A transition is clearly visible over a length scale of $\lambda_i$ about the interface. The asymmetry of the transition region with respect to the atomic interface simply reflects the difference between the mean free paths of the two materials. Within the error bars of our calculation the ratio of the equilibrium values of the currents averaged over the Ir and Pt volumes, $\bar{j}_z^{\rm Ir}/\bar{j}_z^{\rm Pt} = 1.71$, mirrors the ratio $\rho^{800}_{\rm Pt}/\rho^{800}_{\rm Ir}=1.75$ confirming that bulk behaviour is recovered inside the slabs. As shown in \cref{IrPt_j}(b) where the total charge currents in Pt and in Ir are plotted as a function of $z$, the current injected from the leads attains its asymptotic distribution essentially immediately.

In \cref{IrPt_j}(c) we plot streamlines calculated for the charge current in a plane perpendicular to the interface in the Pt$|$Ir bilayer. Streamlines are parallel to the $z$ axis everywhere suggesting no current flow  across the interface in the $x$ direction so each material can be treated as an independent transport channel. A colour map corresponding to the magnitude of the charge current $j=\sqrt{j_x^2+j_z^2}$ is shown in the background for reference. 

\subsection{Effect of different lead materials}
\label{Leads}

\begin{figure*}
\includegraphics[width=\linewidth]{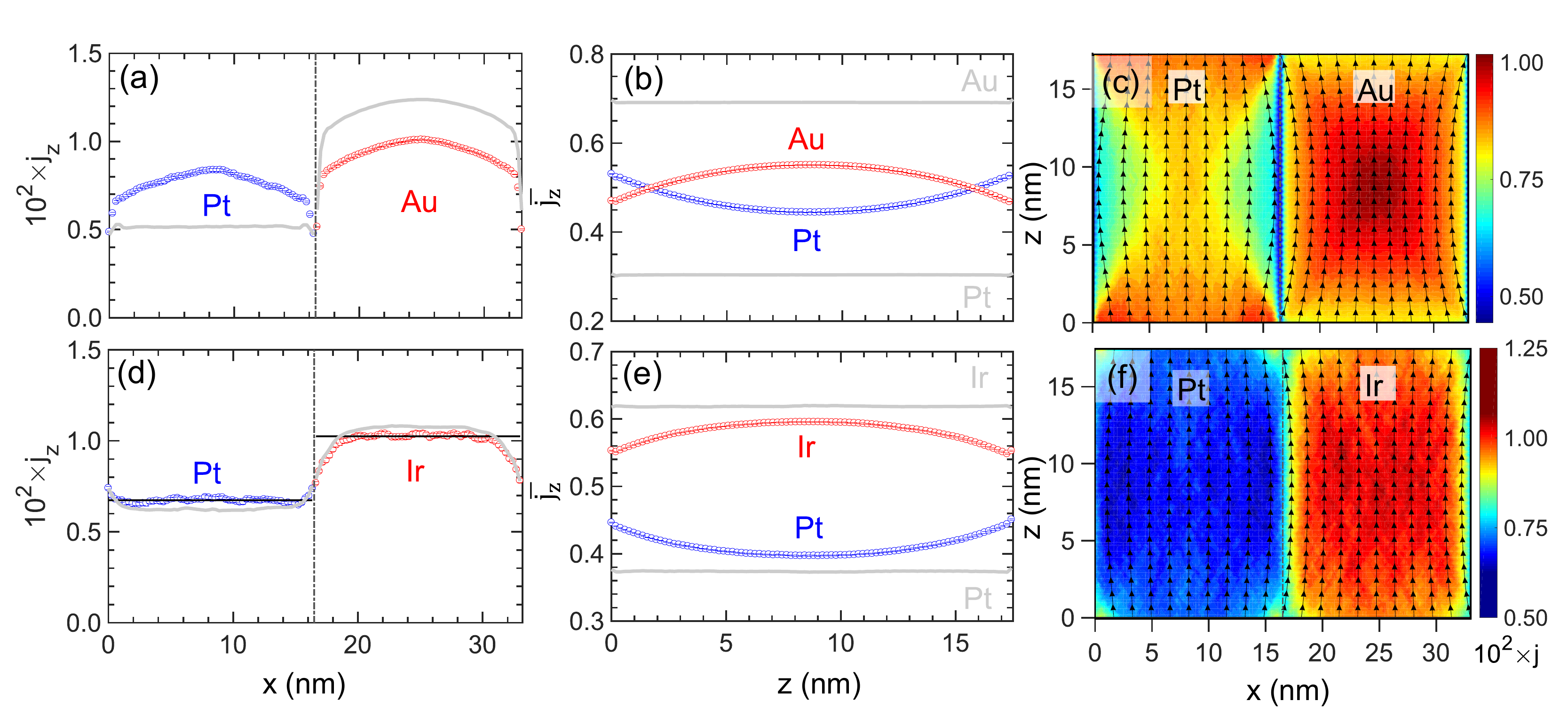}
\caption{Distribution of charge current $\bar{j}_z$ injected from ballistic Pt leads into a 300~K Pt$|$Au multilayer (top) and into a 800~K Pt$|$Ir multilayer (bottom). 
To obtain $\bar{j}_z(x)$ in (a) and (d), $j_z(x,y,z)$ is averaged over $y$ and over the ten central layers in the $z$ direction furthest from the leads. The horizontal black lines in (d) indicate the asymptotic value of $\bar{j}_z$ obtained by averaging $j_z(x,y,z)$ over $y$ and $z$ and $x \in \,$Pt or $x \in \,$Ir omitting atomic layers near the Pt$|$Ir interface where the current varies continuously. 
In (b) and (e), $\bar{j}_z(z)$ is obtained by averaging $j_z(x,y,z)$ over $x$ and $y$ with $x \in \,$Pt and $x \in \,$(Au, Ir) separately. The grey curves indicate the corresponding profiles on injecting from ballistic Au (\cref{AuPt_j}) and Ir (\cref{IrPt_j}) leads into Pt$|$Au and Pt$|$Ir multilayers, respectively. 
(c) and (f) show streamlines of ${\bf j}(x,z)=j_x{\bf i}+j_z{\bf k}$ averaged over $y$ and superimposed on a colour contour of $j(x,z)=\sqrt{j_x^2+j_z^2}$ in the $xz$ plane. Error bars represent the mean deviation over 10 configurations of thermal disorder.}
\label{Jz_multi_Ptlead}
\end{figure*}

Perhaps surprisingly, neither \cref{AuPt_j} nor \cref{IrPt_j} provides any indication of current redistribution from the high resistivity to the low resistivity metal, the phenomenon know as shunting. It transpires that this is because these calculations were carried out using the lower resistivity material of the multilayer pair as the lead material so that there is essentially no contact resistance for the low resistivity channel. Because of the mismatch between their electronic structures, there is a substantial contact resistance between the high conductance lead and the high resistivity material. To lowest order, this contact resistance is the same as the interface resistance between the two materials making up the multilayer. While it is desirable that the results of calculations with the scattering formalism should be independent of the materials used for the leads, this is not always possible in practice because of present memory and computational time constraints. We illustrate this by considering what happens if we use the high resistivity material of the multilayer pair as lead material.   

\subsubsection{Au$|$Pt}

Instead of using Au leads to inject a current into the Au$|$Pt multilayer, we now use Pt. In the Knudsen limit illustrated by the upper panels of \cref{Jz_multi_Ptlead}, the current density does not saturate in the $x$ direction in either Pt or Au (\cref{Jz_multi_Ptlead}(a)) nor does it saturate in the $z$ direction in either Pt or Au for the lengths of scattering region considered here (\cref{Jz_multi_Ptlead}(b)). Compared to the results with Au leads, the current density in Pt is higher whereas that in Au is lower. Because the current does not saturate in the $z$ direction in \cref{Jz_multi_Ptlead}(b), the $\bar{j}_z(z)$ plotted in \cref{Jz_multi_Ptlead}(a) was obtained by averaging $j_z(x, y, z)$ over $y$ and over the ten central layers in the $z$ direction (layers 41 to 50 out of a total of 90). As a function of $x$, the anomalous dip at the interface in the current profile plotted in \Cref{Jz_multi_Ptlead}(a) is much larger than in \cref{AuPt_j}(a) with Au leads. 
 
 In \cref{Jz_multi_Ptlead}(b), the injection of charge carriers from Pt leads into Au (red symbols) is much lower than the injection from Au leads into Au (upper grey line). This can be attributed to the existence of an interface resistance between the ballistic Pt lead and diffusive Au. The converse applies for the injection into diffusive Pt from a ballistic Pt lead which is now higher (blue symbols) than when Au leads were used (lower grey line). Shunting tries to achieve the asymptotic situation where $\rho_{\rm Pt}\bar{j}^{\rm Pt}_z=\rho_{\rm Au}\bar{j}^{\rm Au}_z$ by diverting current from Pt to Au. We see this happening in \cref{Jz_multi_Ptlead}(b) with $\bar{j}^{\rm Au}_z(z)$ increasing and $\bar{j}^{\rm Pt}_z(z)$ decreasing towards the centre of the scattering region. With our present computational resources, the number of layers in the $z$ direction required to achieve asymptotic behaviour is not tractable for the supercell sizes considered here. 
 
 The streamlines plotted in \cref{Jz_multi_Ptlead}(c) now curve towards the Au layers of the multilayer indicating the flow of current across the interface from Pt into Au. Charge transport in the non-asymptotic case studied here is not amenable to description using a simple parallel resistance model following Ohm's law.  It is worthwhile noting that for nanoscale experiments using heterostructures composed of metals with long mean free paths like Cu, asymptotic current distributions cannot be guaranteed for short lengths and the application of semiclassical models may be contentious.

\subsubsection{Pt$|$Ir}

We now look at what happens when Pt leads are used in the classical limit for the 800~K Pt$|$Ir multilayer  illustrated in the lower panes of \cref{Jz_multi_Ptlead}. Comparing \cref{Jz_multi_Ptlead} (a) and (d), we see a striking difference between the classical and Knudsen limits near the interface. In \cref{Jz_multi_Ptlead}(d)  $j_z(x)$ varies gradually across the interface essentially interpolating between the saturated values calculated previously using Ir leads (indicated in grey) with the ratio of the mean values of the current density in each slab (black lines) $\bar{j}_z^{\rm Ir}/\bar{j}_z^{\rm Pt} = 1.52$ falling short of the ratio $\rho^{800}_{\rm Pt}/\rho^{800}_{\rm Ir}=1.71$. As we saw for Pt$|$Au, shunting of the charge current is seen in \cref{Jz_multi_Ptlead}(e) to vary along the transport direction $z$  but the crossover seen near the leads for Pt$|$Au is now absent. And just as we found for Pt$|$Au, the reduction in the current injected into Ir from Pt leads can be attributed to the interface resistance between ballistic Pt and diffusive Ir being larger than the negligible resistance between ballistic Pt and diffusive Pt. Streamlines of the current vector in \cref{Jz_multi_Ptlead}(f) now curve into the Ir layer of the multilayer indicating a net flow from Pt into Ir.
 
The above examples show that longer scattering regions are needed in order to realize the situation where the currents reach their asymptotic distributions that are independent of the choice of lead material. This will become increasingly difficult as the temperature is lowered. The layers considered in this study are comparable in thickness to those use in many experiments in the field of spintronics where Pt layers are typically 10-20~nm thick and current distributions are expected to be asymptotic because of the longer lengths of samples. However, by using the high conductivity material in a multilayer as lead material, we showed that it is possible to probe the asymptotic state.  

\section{Discussion}
\label{discussion}

We have presented a scheme to study spin and charge currents in nontrivial nanostructures containing surfaces and interfaces that builds upon an extremely efficient fully relativistic quantum mechanical scattering formalism \cite{Starikov:prl10, Starikov:prb18} and illustrated it with a study of charge transport in thin films and multilayers of nonmagnetic materials. The specific examples that we considered \textit{viz.} Pt$|$Ir and Pt$|$Au multilayers as well as free-standing thin films of Pt and Au are illustrative of transport regimes where the mean free path $\lambda$ is either much larger than the thickness $d$ of individual layers or much smaller. The ratio $\lambda/d$ is the Knudsen number (Kn) that is well known from fluid physics. As pointed out early on by Fuchs \cite{Fuchs:pcps38}, Sondheimer \cite{Sondheimer:ap52} and others \cite{Lucas:jap65}, it plays a crucial role in determining how currents are distributed near a surface where diffusive scattering leads to a suppression of the current in a thin film as confirmed by numerous experimental studies as well as by our calculations for Au and Pt. The same current suppression is apparent in just the Au layer of a Au$|$Pt multilayer for which $\rm Kn > 1$ but is much smaller in the Pt layer leading to a current density that varies nonmonotonically when we pass from the centre of the Pt layer through the interface to the centre of the Au layer. For a Pt$|$Ir multilayer at $800\,$K for which $\rm Kn < 1$ there is a smooth and continuous variation of the current density  through the interface. 

Although the need to more accurately describe current distributions in metallic multilayers, including transient shunting effects, is widely recognized in order to interpret spin-transport experiments \cite{Ando:jap11, Liu:arXiv11, Niimi:prl11, Morota:prb11}, there has been little progress in devising improved methods of doing so \cite{Stejskal:prb20}. At the same time, in the semiconductor world, there is a growing need to describe electron transport in wires whose size is being constantly reduced \cite{Josell:armr09, Gall:jap16} in order to identify improved interconnect materials.
 While we performed calculations for systems of $\mathcal{O}(10^4)$ atoms, accessing the asymptotic limit at room temperature given by $\rm Kn>1$ for high conducting materials such as Cu with a long mean free path requires calculations on larger systems $\sim \mathcal{O}(10^5)$ atoms. This is currently only limited by computer memory and computational time which will be met by the next generation of computers. In conclusion, our fully resolved current scheme makes it possible to accurately predict electronic transport in the complex geometries frequently encountered in modern microelectronics without introducing empirical parameters.

\section{Acknowledgements}

This work was financially supported by the ``Nederlandse Organisatie voor Wetenschappelijk Onderzoek'' (NWO) through the research programme of the former ``Stichting voor Fundamenteel Onderzoek der Materie,'' (NWO-I, formerly FOM) and through the use of supercomputer facilities of NWO ``Exacte Wetenschappen'' (Physical Sciences). R.S.N acknowledges funding from the Shell-NWO/FOM ``Computational Sciences for Energy Research (CSER)'' PhD program, project number 15CSER12 and is grateful to Max Rang for help with testing the code.


\begin{thebibliography}{85}
\bibitem [{Thi()}]{ThinFilmTransport}%
  \BibitemOpen
  \href@noop {} {}\bibinfo {note} {{Most studies of electronic transport
  through thin films (or nanowires) based on realistic electronic structures
  have been in the ballistic regime \cite{Schep:prl95, Datta:95, Schep:prb98}.
  Most have been for Cu \cite{Zhou:jpcm08, Timoshevskii:jap08, Ke:prb09,
  Feldman:pssb10, Kharche:jpcm11, Roberts:iitc15, Jones:prb15,
  Sanchez-Soares:prb16} but calculations for Ag \cite{Feldman:pssb10,
  Kharche:jpcm11}, and Pt, Rh, Ir and Pd \cite {Lanzillo:jap17} thin films have
  also been reported. A simple but effective ``brute force'' approach to
  modelling the diffusive regime \cite{LiuY:prb11, LiuY:prb15} has to the best
  of our knowledge only been applied to the study of Cu thin films
  \cite{Zhao:jap11, Zhou:prb18} }}\BibitemShut {NoStop}%
\bibitem [{foo({\natexlab{b}})}]{footnote1}%
  \BibitemOpen
  \bibinfo {note} {{We could calculate the
  conductivity entirely from first principles; see \cite{Savrasov:prb96b,
  LiuY:prb15} but because the results are not always in as good agreement with
  experiment as in \cite{LiuY:prb15}, it is more expedient to choose the
  thermal disorder to reproduce the experimental resistivity (and
  magnetization) as in \cite{LiuY:prb15, WangL:prl16, Starikov:prb18,
  Wesselink:prb19, Gupta:prl20} }}\BibitemShut {NoStop}%
\bibitem [{rho()}]{rho:Au300K}%
  \BibitemOpen
  \href@noop {} {}\bibinfo {note} {{The value of root-mean-square displacement
  $\Delta$ chosen with a view to reproducing the room temperature resistivity
  of pure Au turned out to yield a value of $\rho_{\rm Au}=2.6 \pm 0.07 \, \mu
  \Omega \,$cm corresponding to $T\sim 335 \,$K. Since this has no consequences
  for the present study, it was not corrected }}\BibitemShut {NoStop}%
\bibitem [{xya()}]{xyaveraging}%
  \BibitemOpen
  \href@noop {} {}\bibinfo {note} {{Our interatomic currents are normalized to
  the applied bias between the left and right leads. After interpolating these
  onto a regular mesh, the currents are normalized to the conductance in the
  $z$ direction. When summed over $x$ and $y$, they are therefore dimensionless
  and equal to unity at any position $z$ to satisfy charge current
  conservation. $x$ and $y$ averaging is simply given by adding all the
  currents in the $xy$ plane }}\BibitemShut {NoStop}%
\end{thebibliography}
\end{document}